\documentclass[a4paper]{jpconf}
\usepackage{graphicx}
\usepackage{amsmath,amsthm,amsfonts}

\begin{document}
\title{Quantum Gowdy model within the new loop quantum cosmology improved dynamics}

\author{M Mart\'in-Benito${}^{1,2}$, L J Garay${}^{3,2}$,  G A Mena Marug\'an${}^{2}$}

\address{
${}^{1}$MPI f\"ur Gravitational Physics, Albert Einstein Institute,
Am M\"uhlenberg 1, D-14476 Potsdam, Germany.\\
${}^{2}$Instituto de Estructura de la Materia, CSIC,
Serrano 121, 28006 Madrid, Spain.\\
${}^{3}$Departamento de F\'{i}sica Te\'{o}rica II, Universidad
Complutense de Madrid, 28040 Madrid, Spain.}

\ead{mercedes@aei.mpg.de, luis.garay@fis.ucm.es, mena@iem.cfmac.csic.es}

\begin{abstract}
The linearly polarized Gowdy $T^3$ model can be regarded as compact Bianchi I cosmologies with
inhomogeneous modes allowed to travel in one direction. We study a hybrid quantization of this model
that combines the loop quantization of the Bianchi I background, adopting the improved dynamics
scheme put forward by Ashtekar and Wilson-Ewing, with a Fock quantization for the inhomogeneities.
The Hamiltonian constraint operator provides a resolution of the cosmological singularity and
superselects separable sectors. We analyze the complicated structure of these sectors. In any of
them the Hamiltonian constraint provides an evolution equation with respect to the volume of the
associated Bianchi I universe, with a well posed
initial value problem. This fact allows us to construct the Hilbert space of physical states and 
to show that we recover the standard quantum field theory for the inhomogeneities.
\end{abstract}

\section{Introduction}
The Gowdy cosmologies represent a suitable arena to further develop loop quantum cosmology (LQC) \cite{lqc} and include
inhomogeneous settings, since they provide the simplest inhomogeneous cosmological spacetimes \cite{gowd}.
Here, we analyze the particular case of the Gowdy model with three-torus topology and linear polarization.
The homogeneous sector of this model (given by the degrees of freedom that parametrize the subset of homogeneous solutions) coincides with the Bianchi I model with three-torus topology. In turn, the inhomogeneities describe a content of linearly polarized gravitational waves traveling in a single direction.
In previous studies we  carried out the quantization of this system by means of a hybrid procedure \cite{hybrid1,hybrid2}.
More specifically, we applied a Fock quantization for the inhomogeneous sector, in order to deal
with the field complexity, while for the homogeneous sector we employed the quantization of
the Bianchi I model performed in LQC \cite{mmp}, with the main aim of achieving a quantum resolution of
the cosmological singularity.

The quantization of the Bianchi I model in LQC is subject to several ambiguities.  One of them
concerns the representation of the curvature tensor of the homogeneous connection in the quantum
theory. Different definitions of this object lead to different schemes of quantization. In this
respect,  Ashtekar and Wilson-Ewing have put forward the quantization of the Bianchi I model within
a new {\sl improved dynamics} scheme \cite{awe}, which seems to be physically more appropriate than the
scheme employed in \cite{hybrid1,hybrid2,mmp}.
In this note we summarize the hybrid quantization of the Gowdy model that results from adopting this
new {improved dynamics} scheme in the representation of the homogeneous sector. We focus on the new
features that the use of this scheme introduces, such as the complicated structure of the
resulting sectors of superselection (aspect not analyzed in \cite{awe}) and the interpretation of
the quantum Hamiltonian constraint as an evolution equation with respect to the volume of the
associated Bianchi I universe. We will discus that this interpretation is valid inasmuch as the
corresponding initial value problem is well posed: physical solutions are completely determined by a
countable set of data given on an initial section of non-vanishing Bianchi I volume. Moreover, this
fact allows us to identify the physical Hilbert space with the Hilbert space of initial data. For the
details of this work we refer to \cite{hybrid3}.

\section{Quantization of the model}

The linearly polarized Gowdy $T^3$ cosmologies are globally hyperbolic spacetimes with three-torus spatial topology and with two axial and hypersurface orthogonal Killing vector fields $\partial_{\sigma}$ and $\partial_{\delta}$ \cite{gowd}. We use global coordinates adapted to the symmetries $\lbrace t, \theta, \sigma, \delta \rbrace$, with
$\theta,\sigma, \delta \in S^{1}$. Then the spatial dependence occurs only in $\theta$, and we can expand
the field-like variables in Fourier series in that coordinate. After performing a partial gauge
fixing
\cite{hybrid2,hybrid3}, the reduced phase space is described
by two pairs of canonically conjugate point-particle variables (they do not depend
on $\theta$) and by one field $\xi$, together with its canonical momentum $P_{\xi}$. Two global constraints remain in the model,
the spatial average of the $\theta$-momentum constraint (which generates translations in $S^1$) and
the spatial average of the densitized Hamiltonian constraint.

We call inhomogeneous sector the set of degrees of freedom given by the nonzero Fourier modes of
$\xi$ and those of $P_{\xi}$. To describe this sector we introduce creation and
annihilation-like variables, denoted by $\{a^\ast_m,a_m\}$ ($m\in\mathbb{Z}-\{0\}$), defined as
if $\xi$ and $P_{\xi}$ were a canonical pair for a free massless scalar field. In the
quantum theory we promote them to annihilation and creation operators, with
$[\hat{a}_m,\hat{a}^\dagger_m]=1$. $\mathcal F$ will denote the corresponding Fock space.
The generator of translations on the circle only involves this
inhomogeneous sector. It is not difficult to obtain its quantum counterpart in terms of the basic
operators $\hat{a}_m$ and $\hat{a}^\dagger_m$.
The states that verify this constraint form a Fock
subspace $\mathcal F_p\subset\mathcal F$ \cite{hybrid2,hybrid3}.

In turn, the homogeneous sector is formed by the rest of degrees of freedom: the two
point-particle variables and the zero mode of $\xi$, together with its momentum. This sector
coincides with the phase space of the Bianchi I model. We describe it in the Ashtekar-Barbero
formalism.
Using a diagonal gauge, the nontrivial
components of the densitized triad are $p_{i}/4\pi^{2}$, with $i=\theta, \sigma,
\delta$, and $c_{i}/2{\pi}$ are those of the $su(2)$ connection.
They satisfy $\left\{c_{i},p_{j} \right\}=8\pi G \gamma
\delta_{ij}$, where $\gamma$ is the Immirzi parameter and $G$ is the Newton constant. In order to
represent this sector in the quantum theory we choose the kinematical Hilbert space of the Bianchi I
model constructed in LQC \cite{mmp,awe}, that we call
$\mathcal{H}_{\text{kin}}^{\text{BI}}$. We recall that, on $\mathcal{H}_{\text{kin}}^{\text{BI}}$, the operators
$\hat{p}_{i}$ have a discrete
spectrum equal to the real line. The corresponding eigenstates,
$|p_\theta,p_\sigma,p_\delta\rangle$, form an orthonormal basis (in the
discrete norm) of $\mathcal{H}_{\text{kin}}^{\text{BI}}$. Owing to this
discreteness, there is no well defined operator representing the connection,
but rather its holonomies. They are computed along straight edges in the fiducial directions.
The so-called improved dynamics prescription
states that, when writing the curvature tensor in terms of holonomies, we have to evaluate them
along edges with a certain minimum dynamical (state dependent) length
$\bar\mu_i$. We use the specific improved dynamics
prescription put forward in \cite{awe}: all the elementary operators which represent the matrix
elements of
the holonomies, called $\hat{\mathcal{N}}_{\bar{\mu}_{i}}$, produce a constant
shift in the Bianchi I volume of
the compact spatial section. In order
to simplify the analysis, it is convenient to relabel the basis states
in the form $|v,\lambda_\sigma,\lambda_\delta\rangle$, where $v$ is proportional to the volume and such
that the operators $\hat{\mathcal{N}}_{\pm\bar{\mu}_{i}}$ cause a shift on it equal to $\pm 1$,
and the parameters $\lambda_i$ are
all equally defined in terms of the corresponding parameters $p_i$, and verify
that $v=2\lambda_\theta\lambda_\sigma\lambda_\delta$.

The Hamiltonian constraint of the Gowdy model is formed by the Hamiltonian constraint of the
Bianchi I model plus the coupling term which involves both homogeneous and inhomogeneous sectors.
Out of the basic operators $\hat{p}_i$, $\hat{\mathcal{N}}_{\bar{\mu}_{i}}$, $\hat{a}_m$, and
$\hat{a}_m^\dagger$, we represent it as an operator $\widehat{\mathcal C}$ well defined on a dense
domain of the kinematical Hilbert space $\mathcal{H}_{\text{kin}}^{\text{BI}}\otimes\mathcal F$
\cite{hybrid3}.
We choose a very suitable symmetric factor ordering such that $\widehat{\mathcal C}$ decouples the
states of $\mathcal{H}_{\text{kin}}^{\text{BI}}\otimes\mathcal F$ with support on $v=0$, namely,
the states with vanishing homogeneous volume. Moreover, our operator $\widehat{\mathcal C}$ does
not relate states with different orientation of any of the eigenvalues of the operators
$\hat{p}_i$. Owing to this property we can then restrict the homogeneous sector (also in the
domain of definition of $\widehat{\mathcal C}$) to e.g. the space spanned by the states
$|v>0,\lambda_\sigma>0,\lambda_\delta>0\rangle$. We call the resulting Hilbert space
$\mathcal{H}_{\text{kin}}^{\text{BI},+}$. Remarkably, for this restriction of the homogeneous
sector we do not need to impose any particular
boundary condition or appeal to any parity symmetry. Note that both
$\mathcal{H}_{\text{kin}}^{\text{BI}}$ and $\mathcal{H}_{\text{kin}}^{\text{BI},+}$ are
non-separable Hilbert spaces, feature which is not desirable for a physical theory. This problem is overcome
by the very action of the Hamiltonian constraint operator. Indeed, $\widehat{\mathcal C}$ defined on
(a dense domain in)
$\mathcal{H}_{\text{kin}}^{\text{BI},+}\otimes\mathcal F$
turns out to leave invariant some separable Hilbert subspaces that provide superselection sectors.
Specifically, the homogeneous sector of these superselection sectors is the space spanned by the
states
$|v=\varepsilon+4n,
\lambda_\sigma=\lambda^{\star}_{\sigma}\omega_{\varepsilon},\lambda_\delta=\lambda^{\star}_{\delta}
\omega_ { \varepsilon } \rangle$. Here $\varepsilon\in(0,4]$ and $n\in\mathbb{N}$; therefore
$v$ takes support on semilattices of constant step equal to $4$ starting in a minimum non-vanishing
value $\varepsilon$. In addition, $\lambda^{\star}_{a}$ ($a=\sigma,\delta$) is some positive real
number and
$\omega_{\varepsilon}$ runs over the subset of $\mathbb{R}^{+}$ given by
\begin{equation}
\left\{\left(\frac{\varepsilon-2}{\varepsilon}\right)^{z}
\prod_{k}\left(\frac{\varepsilon+2m_k}{\varepsilon+2n_{k}} \right)^{p_{k}}\right\},\nonumber
\end{equation}
where $m_{k},n_{k},p_{k}\in \mathbb{N}$, and $z \in \mathbb{Z}$ when $\varepsilon>2$, while $z=0$
otherwise. One can check that indeed this set is dense in $\mathbb{R}^{+}$ and countable
\cite{hybrid3}. Therefore, any of these sectors provide separable Hilbert spaces
contained in $\mathcal{H}_{\text{kin}}^{\text{BI}}$. We denote them as $\mathcal
H_{\varepsilon,\lambda_\sigma^\star,\lambda_\delta^\star}=\mathcal H_\varepsilon\otimes\mathcal
H_{\lambda_\sigma^\star}\otimes\mathcal H_{\lambda_\delta^\star}$.
Note that the fact that the states with support on $v=0$ 
are removed means that there is no analog in our quantum theory of the cosmological
singularities, classically requiring that some $p_i$ vanish. We thus solve these
singularities already at the kinematical level in a very simple way.

We then restrict the study to any of these superselection sectors and define the homogeneous terms
of the Hamiltonian constraint operator $\widehat{\mathcal C}$ on the span of the states
$|\varepsilon+4n,\lambda^{\star}_{\sigma}\omega_{\varepsilon},\lambda^{\star}_{\delta}\omega_{
\varepsilon}\rangle$.
The
Hamiltonian constraint turns out to provide a difference equation in the parameter $v$
\cite{hybrid3}. Therefore, if we regard
$v$ as an
internal time, the constraint can be interpreted as an evolution equation in it.
The condition that this constraint imposes is very complicated and the solutions can be obtained
only formally.
This is in part owing to the fact that one of the inhomogeneous terms creates in every step of the
evolution an infinite number of particles (for a detailed discussion see
\cite{hybrid1,hybrid2,hybrid3}). Nonetheless, thanks to the precise
structure of the superselection sectors  for the variables $\lambda_a$ we can argue that a set of
countable data given on the initial section $v=\varepsilon$ completely determines the
physical
solutions. This issue is non-trivial. Actually, both the separability of the support of
$\lambda_a$ and the fact that it is dense in $\mathbb{R}^+$ have been essential to show it (the
details can be found in \cite{hybrid4}). At the end of the day, the initial value problem is well posed and it
is indeed feasible to make the above interpretation of the constraint as an evolution equation in
$v$.

We can thus
identify solutions with initial data. Furthermore, we can
construct a(n over) complete set of observables
acting on the space of initial data and impose reality conditions on them \cite{hybrid3,hybrid4}.
This determines a unique inner product that endows the space of initial data with
a
Hilbert structure. Then, we finally impose the $S^1$ symmetry and identify the resulting Hilbert
space with the physical Hilbert space. The result is that the physical Hilbert space is
$H_{\lambda_\sigma^\star}\otimes\mathcal
H_{\lambda_\delta^\star}\otimes\mathcal F_p$, with $H_{\lambda_\sigma^\star}\otimes\mathcal
H_{\lambda_\delta^\star}$ being the physical Hilbert space of the Bianchi I model in LQC \cite{hybrid4}.

\section{Conclusions}
This work summarizes the quantization in the framework of LQC of the Gowdy $T^3$ model with linear polarization,
which is an inhomogeneous cosmological spacetime with spatial dependence in a single direction
$\theta$. In order to perform it, first, at the classical level, we fix the gauge partially. 
The reduced system is subject to two global constraints: the spatial average of the
$\theta$-momentum constraint and the spatial average
of the densitized Hamiltonian constraint. These constraints are imposed \`a la Dirac in the quantum
theory. Moreover, they are represented by means of a hybrid approach, which
combines the polymeric techniques of LQC with those of the Fock quantization. The former ones,
applied in the representation of the homogeneous sector of the system, allow us to resolve the
cosmological singularity in the quantum theory, while the Fock quantization for the inhomogeneities
successfully deal with the field complexity.

Employing this hybrid approach we obtain well defined operators for the constraints. For the
Hamiltonian constraint operator, we choose a very convenient symmetric factor ordering, such that it
allows us to restrict the study of the homogeneous sector to suitable, separable superselection sectors.
In any of them, the Hamiltonian constraint provides a difference equation in the volume of
the homogeneous sector of the Gowdy model, which coincides with the Bianchi I model.
We have interpreted this volume as
an internal time and the Hamiltonian constraint as an evolution equation in that time. We have
argued that this interpretation is valid, inasmuch as the corresponding initial value problem is
well posed. In fact, a countable dense set of initial data completely specifies the physical
solutions, and then we can characterize the physical Hilbert space as a Hilbert space of
initial data. The physical inner product is fixed by demanding reality conditions on a complete set
of observables and imposing the $S^1$ symmetry.
We conclude that the physical Hilbert space is the physical Hilbert space of the Bianchi I model in
LQC tensor product a Fock space $\mathcal F_p$.

It is important to note that, if the model is totally
deparametrized and quantized as in standard
quantum field theory, it admits a unique Fock quantization with unitary dynamics and such that the
vacuum is invariant under $S^1$ translations \cite{ccm,ccmv1,ccmv2}.
The physical Hilbert space for the inhomogeneities, namely $\mathcal F_p$, is actually
equivalent to that obtained for them in the standard quantization of the deparametrized model. Therefore we
recover the standard description of the inhomogeneities, which can be seen as degrees of
freedom propagating over the polymerically quantized Bianchi I background. This result, although
expected, is not trivial since homogeneous and inhomogeneous sectors have been quantized with very
different methods and they are coupled in a very involved way by the Hamiltonian constraint.

\ack This work was in part supported by the Spanish MICINN Project No. FIS2008-06078-C03-03 and
the Consolider-Ingenio Program CPAN No. CSD2007-00042. M. M-B has been
supported by CSIC and the European Social Fund under the grant I3P-BPD2006.

\end{document}